\documentclass[a4paper,12pt]{article}
\usepackage{amsfonts}
\usepackage{graphicx}
\usepackage{amsmath}
\usepackage{hyperref}

\begin{document}

\begin{center}
{\huge On a particular form of a symmetric P\"{o}schl-Teller potential}

\textbf{Andrei Smirnov\footnote{%
email: smirnov@ufs.br, smirnov.globe@gmail.com}, Antonio Jorge Dantas Farias
Jr.\footnote{%
email: a.jorgedantas@ufs.br, a.jorgedantas@gmail.com}}

\textit{Universidade Federal de Sergipe}

Abstract
\end{center}

{\small We show that solutions of the Schr\"{o}dinger equation with a
symmetric P\"{o}schl-Teller potential of a particular form can be expressed
in terms of a closed combination (not series) of trigonometric functions.
Using some properties of the eigenfunctions of the Schr\"{o}dinger equation
and their inner product we determine a new exact representation of the
hypergeometric function with certain values of parameters in terms of a
closed combination of trigonometric functions. We also obtain new results in
an explicit closed form for integrals with the hypergeometric function and
with the specific combination of trigonometric functions.}

\section{Solutions in terms of hypergeometric function}

There are well known solutions of the Schr\"{o}dinger equation 
\begin{equation}
\left[ -\frac{d^{2}}{dx^{2}}+U\left( x\right) \right] \psi =E\psi
\label{r01}
\end{equation}%
with the P\"{o}schl-Teller potential%
\begin{equation}
U\left( x\right) =\alpha ^{2}\left[ \frac{\kappa \left( \kappa -1\right) }{%
\sin ^{2}\left( \alpha x\right) }+\frac{\lambda \left( \lambda -1\right) }{%
\cos ^{2}\left( \alpha x\right) }\right] ;\ \kappa >1,\ \lambda >1
\label{r02}
\end{equation}%
in the interval $\left( 0,\frac{\pi }{2\alpha }\right) $, which are
expressed in terms of hypergeometric function $\left. _{2}F_{1}\right.
\left( \alpha ,\beta ;\gamma ;x\right) =\sum_{n=0}^{\infty }\frac{\left(
\alpha \right) _{n}\left( \beta \right) _{n}}{\left( \gamma \right) _{n}}%
\frac{x^{n}}{n!}$:%
\begin{equation}
\psi _{n}\left( x\right) =A_{n}\sin ^{\kappa }\left( \alpha x\right) \cos
^{\lambda }\left( \alpha x\right) \left. _{2}F_{1}\right. \left( -n,n+\kappa
+\lambda ;\kappa +\frac{1}{2};\sin ^{2}\left( \alpha x\right) \right) ,\
n\geq 0\ ,  \label{r03}
\end{equation}%
$A_{n}$ is a normalization factor, with energy spectrum 
\begin{equation}
E_{n}=\alpha ^{2}\left( 2n+\kappa +\lambda \right) ^{2}\ .  \label{r04}
\end{equation}%
These solutions are described, for example, in Ref. \cite{f99}, problem 38
and presented in Ref. \cite{bg90}, Appendix A, problem A.I.6. We consider a
particular form of the P\"{o}schl-Teller potential with values of the
parameters $\kappa =\lambda =2$ and obtain solutions in terms of usual
trigonometric functions in a closed form (not series).

\section{Solutions generated by Darboux transformation method}

In order to obtain the solutions in terms of trigonometric functions we
apply the Darboux transformation (DT) method \cite{matveev-salle}, \cite%
{bs95}, \cite{bs97} to the Schr\"{o}dinger equation with the infinite
rectangular well potential in the same interval $\left( 0,\frac{\pi }{%
2\alpha }\right) $. Then the initial operator of the DT method is the
Hamiltonian%
\begin{equation}
\widehat{H}_{0}=-\frac{d^{2}}{dx^{2}},\ x\in \left( 0,\frac{\pi }{2\alpha }%
\right) ;\ \widehat{H}_{0}\varphi _{k}=\varepsilon _{k}\varphi _{k}\ .
\label{r10}
\end{equation}%
Its eigenfunctions and energy spectrum are given by%
\begin{equation}
\varphi _{k}\left( x\right) =\sqrt{\frac{4\alpha }{\pi }}\sin \left( 2\alpha
kx\right) ,\ \varepsilon _{k}=4\alpha ^{2}k^{2},\ k\geq 1\ .  \label{r09}
\end{equation}%
We note that eigenfunctions $\varphi _{k}$ in Eq. (\ref{r09}) are
orthonormalized. In accordance with the DT method we represent $\widehat{H}%
_{0}$ in the form%
\begin{equation*}
\widehat{H}_{0}=\widehat{L}^{\dagger }\widehat{L}+\omega ^{2}\ ,
\end{equation*}%
where $\omega ^{2}$ is a real constant, $\widehat{L}$ is the intertwining
operator, and $\widehat{L}^{\dagger }$ is its adjoint: 
\begin{equation}
\widehat{L}=\frac{d}{dx}+W\left( x\right) ,\ \widehat{L}^{\dagger }=-\frac{d%
}{dx}+W\left( x\right) \ .  \label{r11}
\end{equation}%
Applying $\widehat{L}$ to both sides of Eq. (\ref{r10}) one has%
\begin{equation*}
\widehat{L}\widehat{H}_{0}\varphi _{k}=\widehat{L}\left( \widehat{L}%
^{\dagger }\widehat{L}+\omega ^{2}\right) \varphi _{k}=\left( \widehat{L}%
\widehat{L}^{\dagger }+\omega ^{2}\right) \widehat{L}\varphi
_{k}=\varepsilon _{k}\widehat{L}\varphi _{k}\ 
\end{equation*}%
that can be written as%
\begin{equation}
\widehat{H}_{1}\chi _{k}=\varepsilon _{k}\chi _{k},\ \widehat{H}_{1}=%
\widehat{L}\widehat{L}^{\dagger }+\omega ^{2},\ \chi _{k}=\widehat{L}\varphi
_{k}\ ,  \label{r12}
\end{equation}%
where $\chi _{k}$ are eigenfunctions of $\widehat{H}_{1}$. The spectrum $%
\varepsilon _{k}$ of $\widehat{H}_{1}$ is the same of $\widehat{H}_{0}$,
except maybe some particular values. Using the relations (\ref{r11}), we
write%
\begin{equation}
\widehat{H}_{1}=-\frac{d^{2}}{dx^{2}}+V_{1}\left( x\right) ,\ V_{1}\left(
x\right) =W^{\prime }+W^{2}+\omega ^{2}  \label{r14}
\end{equation}%
Choosing $W\left( x\right) $, $\omega ^{2}$ as follows%
\begin{equation}
W\left( x\right) =-\frac{d}{dx}\ln \left( \varphi _{1}\right) =-\frac{%
\varphi _{1}^{\prime }}{\varphi _{1}},\ \omega ^{2}=\varepsilon _{1}\ ,
\label{r16}
\end{equation}%
one has for them from Eq. (\ref{r09})%
\begin{equation*}
W\left( x\right) =-2\alpha \cot \left( 2\alpha x\right) ,\ \omega
^{2}=4\alpha ^{2}
\end{equation*}%
and for $V_{1}\left( x\right) $ of Eq. (\ref{r14})%
\begin{equation}
V_{1}\left( x\right) =2\left( \frac{\varphi _{1}^{\prime 2}}{\varphi _{1}^{2}%
}+\varepsilon _{1}^{2}\right) =2\left( 2\alpha \right) ^{2}\left( \cot
^{2}\left( 2\alpha x\right) +1\right) =\frac{8\alpha ^{2}}{\sin ^{2}\left(
2\alpha x\right) }\ .  \label{r15}
\end{equation}%
To obtain the eigenfunctions of $\widehat{H}_{1}$ by means of the DT method
we apply $\widehat{L}$ to the eigenfunctoions $\varphi _{k}$ of the initial
Hamiltonian $\widehat{H}_{0}$ that follows from Eq. (\ref{r12})%
\begin{equation}
\chi _{k}=\widehat{L}\varphi _{k}=\frac{d\varphi _{k}}{dx}+W\left( x\right)
\varphi _{k}  \label{r21}
\end{equation}%
\begin{equation*}
=\sqrt{\frac{4\alpha }{\pi }}2\alpha \left[ k\cos \left( 2\alpha kx\right)
-\cot \left( 2\alpha x\right) \sin \left( 2\alpha kx\right) \right] ,\ k\geq
2\ .
\end{equation*}%
The eigenfunctoions of (\ref{r21}) are not normalized. To normalize them we
use the following properties of an inner product of the eigenfunctions:%
\begin{equation*}
\left( \chi _{i},\chi _{j}\right) =\left( \widehat{L}\varphi _{i},\widehat{L}%
\varphi _{j}\right) =\left( \widehat{L}^{\dagger }\widehat{L}\varphi
_{i},\varphi _{j}\right) =\left( \left( \widehat{H}_{0}-\omega ^{2}\right)
\varphi _{i},\varphi _{j}\right) 
\end{equation*}%
\begin{equation*}
=\left( \left( \varepsilon _{i}-\omega ^{2}\right) \varphi _{i},\varphi
_{j}\right) =\left( \varepsilon _{i}-\omega ^{2}\right) \left( \varphi
_{i},\varphi _{j}\right) =\left( \varepsilon _{i}-\omega ^{2}\right) \delta
_{ij}\ .
\end{equation*}%
Designating the normalized eigenfunctoions by $\widetilde{\chi }_{k}=%
\widetilde{N}_{k}\chi _{k}$, one has%
\begin{equation*}
\left( \widetilde{\chi }_{k},\widetilde{\chi }_{k}\right) =\left( \widetilde{%
N}_{k}\chi _{k},\widetilde{N}_{k}\chi _{k}\right) =\widetilde{N}%
_{k}^{2}\left( \chi _{k},\chi _{k}\right) =\widetilde{N}_{k}^{2}\left(
\varepsilon _{k}-\omega ^{2}\right) =1\ ,
\end{equation*}%
therefore%
\begin{equation*}
\widetilde{N}_{k}=\frac{1}{\sqrt{\varepsilon _{k}-\omega ^{2}}}\ .
\end{equation*}%
For the choice of $\omega ^{2}$ indicated in Eq. (\ref{r16}) and with use of
Eq. (\ref{r09}) one has%
\begin{equation*}
\widetilde{N}_{k}=\frac{1}{\sqrt{\varepsilon _{k}-\varepsilon _{1}}}=\frac{1%
}{2\alpha \sqrt{k^{2}-1}}\ .
\end{equation*}%
Then the normalized eigenfunctions and the energy spectrum are%
\begin{equation}
\widetilde{\chi }_{k}\left( x\right) =N_{k}\left[ k\cos \left( 2\alpha
kx\right) -\cot \left( 2\alpha x\right) \sin \left( 2\alpha kx\right) \right]
,\ N_{k}=\sqrt{\frac{4\alpha }{\pi }}\frac{1}{\sqrt{k^{2}-1}},\ \varepsilon
_{k}=4\alpha ^{2}k^{2},\ k\geq 2\ .  \label{r22}
\end{equation}%
For the ground state energy one has%
\begin{equation}
\varepsilon _{2}=4\alpha ^{2}2^{2}\ .  \label{r24}
\end{equation}%
We note that at the middle of the interval, $x=\frac{\pi }{4\alpha }$:%
\begin{equation}
\widetilde{\chi }_{k}\left( \frac{\pi }{4\alpha }\right) =N_{k}\left[ k\cos
\left( k\frac{\pi }{2}\right) \right] ,\ \widetilde{\chi }_{k}^{\prime
}\left( \frac{\pi }{4\alpha }\right) =N_{k}2\alpha \left[ 1-k^{2}\right]
\sin \left( k\frac{\pi }{2}\right) \ ,  \label{r31}
\end{equation}%
then for even $k=2m$:%
\begin{equation}
\widetilde{\chi }_{2m}\left( \frac{\pi }{4\alpha }\right) =N_{2m}2m\left(
-1\right) ^{m},\ \widetilde{\chi }_{2m}^{\prime }\left( \frac{\pi }{4\alpha }%
\right) =0  \label{r32}
\end{equation}%
and for odd $k=2m+1$:%
\begin{equation}
\widetilde{\chi }_{2m+1}\left( \frac{\pi }{4\alpha }\right) =0,\ \widetilde{%
\chi }_{2m+1}^{\prime }\left( \frac{\pi }{4\alpha }\right) =N_{2m+1}2\alpha %
\left[ 1-\left( 2m+1\right) ^{2}\right] \left( -1\right) ^{m}\ .  \label{r33}
\end{equation}

\section{Correspondence between the solutions}

The potential $V_{1}\left( x\right) $ of Eq. (\ref{r15}) corresponds to the P%
\"{o}schl-Teller potential $U\left( x\right) $ of Eq. (\ref{r02}) with the
values of the parameters $\kappa =\lambda =2$: 
\begin{equation*}
U\left( x\right) =2\alpha ^{2}\left( \frac{1}{\sin ^{2}\left( \alpha
x\right) }+\frac{1}{\cos ^{2}\left( \alpha x\right) }\right) =\frac{2\alpha
^{2}}{\sin ^{2}\left( \alpha x\right) \cos ^{2}\left( \alpha x\right) }=%
\frac{8\alpha ^{2}}{\sin ^{2}\left( 2\alpha x\right) }\ .
\end{equation*}%
The eigenfunctions and the energy in the form of Eqs. (\ref{r03}), (\ref{r04}%
) with $\kappa =\lambda =2$ are%
\begin{eqnarray}
&&\psi _{n}\left( x\right) =A_{n}\sin ^{2}\left( \alpha x\right) \cos
^{2}\left( \alpha x\right) \left. _{2}F_{1}\right. \left( -n,n+4;\frac{5}{2}%
;\sin ^{2}\left( \alpha x\right) \right) \ ,  \label{r25} \\
&&E_{n}=\alpha ^{2}\left( 2n+4\right) ^{2}=4\alpha ^{2}\left( n+2\right)
^{2},\ n\geq 0\ .  \notag
\end{eqnarray}%
For the ground state energy one has%
\begin{equation}
E_{0}=4\alpha ^{2}2^{2}\ .  \label{r26}
\end{equation}%
Taking into account Eqs. (\ref{r24}), (\ref{r26}) and Eqs. (\ref{r22}), (\ref%
{r25}), the following correspondence takes place for energies and the
normalized eigenfunctions:%
\begin{equation}
E_{n}=\varepsilon _{n+2},\ \psi _{n}\left( x\right) =\widetilde{\chi }%
_{n+2}\left( x\right) ,\ n\geq 0\ .  \label{r27}
\end{equation}%
where due to Eq. (\ref{r22}) $\widetilde{\chi }_{n+2}\left( x\right) $ is 
\begin{eqnarray}
&&\widetilde{\chi }_{n+2}\left( x\right) =N_{n+2}\left[ \left( n+2\right)
\cos \left( 2\left( n+2\right) \alpha x\right) -\cot \left( 2\alpha x\right)
\sin \left( 2\left( n+2\right) \alpha x\right) \right] ,\   \label{r23} \\
&&N_{n+2}=\sqrt{\frac{4\alpha }{\pi }}\frac{1}{\sqrt{\left( n+2\right) ^{2}-1%
}},\ n\geq 0\ .  \notag
\end{eqnarray}%
We also note that at the middle of the interval, $x=\frac{\pi }{4\alpha }$:%
\begin{equation}
\psi _{n}\left( \frac{\pi }{4\alpha }\right) =A_{n}\frac{1}{4}\left.
_{2}F_{1}\right. \left( -n,n+4;\frac{5}{2};\frac{1}{2}\right) \ ,
\label{r28}
\end{equation}%
\begin{equation}
\psi _{n}^{\prime }\left( \frac{\pi }{4\alpha }\right) =A_{n}\frac{\alpha }{%
10}\left( -n\right) \left( n+4\right) \left. _{2}F_{1}\right. \left(
-n+1,n+5;\frac{7}{2};\frac{1}{2}\right) \ ,  \label{r29}
\end{equation}%
where the relation%
\begin{equation*}
\frac{d}{dx}\left. _{2}F_{1}\right. \left( \alpha ,\beta ;\gamma ;x\right) =%
\frac{\alpha \beta }{\gamma }\left. _{2}F_{1}\right. \left( \alpha +1,\beta
+1;\gamma +1;x\right)
\end{equation*}%
was used.

\section{Some new properties of the hypergeometric function}

Now we are at position to establish an expression of the hypergeometric
function of the form of Eq. (\ref{r25}) in terms of a composition of
trigonometric functions. We write the normalization factor $A_{n}$ of Eq. (%
\ref{r25}) as follows%
\begin{equation}
A_{n}=C_{n}^{-1}N_{n+2},\ N_{n+2}=\sqrt{\frac{4\alpha }{\pi }}\frac{1}{\sqrt{%
\left( n+2\right) ^{2}-1}}\ .  \label{r30}
\end{equation}%
Then from Eq. (\ref{r27}) and Eqs. (\ref{r25}), (\ref{r23}) one has 
\begin{equation}
\left. _{2}F_{1}\right. \left( -n,n+4;\frac{5}{2};\sin ^{2}\left( \alpha
x\right) \right) =4C_{n}\frac{\left[ \left( n+2\right) \cos \left( 2\alpha
\left( n+2\right) x\right) -\cot \left( 2\alpha x\right) \sin \left( 2\alpha
\left( n+2\right) x\right) \right] }{\sin ^{2}\left( 2\alpha x\right) }\ .
\label{r34}
\end{equation}%
To determine $C_{n}$ in Eq. (\ref{r34}) we write Eq. (\ref{r27}) for the
eigenfunctions and their derivatives at the point $x=\frac{\pi }{4\alpha }$.
With use of Eq. (\ref{r31}) one has%
\begin{equation}
\widetilde{\chi }_{n+2}\left( \frac{\pi }{4\alpha }\right) =-N_{n+2}\left[
\left( n+2\right) \cos \left( n\frac{\pi }{2}\right) \right] ,\ \widetilde{%
\chi }_{n+2}^{\prime }\left( \frac{\pi }{4\alpha }\right) =N_{n+2}2\alpha %
\left[ \left( n+2\right) ^{2}-1\right] \sin \left( n\frac{\pi }{2}\right) \ .
\label{r35}
\end{equation}%
Then from Eqs. (\ref{r28}), (\ref{r29}), (\ref{r35}) for even $n=2m$:%
\begin{equation}
\frac{1}{4}\left. _{2}F_{1}\right. \left( -2m,2m+4;\frac{5}{2};\frac{1}{2}%
\right) =C_{2m}2\left( m+1\right) \left( -1\right) ^{m+1}\ ,  \label{r36}
\end{equation}%
therefore%
\begin{equation}
C_{2m}=\frac{\left( -1\right) ^{m+1}}{8\left( m+1\right) }\left.
_{2}F_{1}\right. \left( -2m,2m+4;\frac{5}{2};\frac{1}{2}\right) \ ;
\label{r37}
\end{equation}%
for odd $n=2m+1$:%
\begin{equation}
\frac{\alpha }{10}\left( -\left( 2m+1\right) \right) \left( 2m+5\right)
\left. _{2}F_{1}\right. \left( -2m,2m+6;\frac{7}{2};\frac{1}{2}\right)
=C_{2m+1}2\alpha \left[ \left( 2m+3\right) ^{2}-1\right] \left( -1\right)
^{m}\ ,  \label{r38}
\end{equation}%
therefore%
\begin{equation}
C_{2m+1}=\frac{\left( -1\right) ^{m+1}}{20}\frac{\left( 2m+1\right) \left(
2m+5\right) }{4\left( m+1\right) \left( m+2\right) }\left. _{2}F_{1}\right.
\left( -2m,2m+6;\frac{7}{2};\frac{1}{2}\right) \ .  \label{r39}
\end{equation}%
We also note that due to the properties of the solutions (\ref{r32}), (\ref%
{r33}) one can conclude that%
\begin{equation}
\left. _{2}F_{1}\right. \left( -\left( 2m+1\right) ,2m+5;\frac{5}{2};\frac{1%
}{2}\right) =0,\ \left. _{2}F_{1}\right. \left( -2m+1,2m+5;\frac{7}{2};\frac{%
1}{2}\right) =0,\ m=0,1,2,...\ .  \label{r40}
\end{equation}%
Subsituting Eqs. (\ref{r37}), (\ref{r39}) into Eq. (\ref{r34}), one obtains
explicitly%
\begin{eqnarray}
&&\frac{\left. _{2}F_{1}\right. \left( -2m,2m+4;\frac{5}{2};\sin ^{2}\left(
\alpha x\right) \right) }{\left. _{2}F_{1}\right. \left( -2m,2m+4;\frac{5}{2}%
;\frac{1}{2}\right) }=\frac{\left( -1\right) ^{m+1}}{2\left( m+1\right) }
\label{r41} \\
&&\times \frac{\left[ \left( 2m+2\right) \cos \left( 2\alpha \left(
2m+2\right) x\right) -\cot \left( 2\alpha x\right) \sin \left( 2\alpha
\left( 2m+2\right) x\right) \right] }{\sin ^{2}\left( 2\alpha x\right) } 
\notag
\end{eqnarray}%
and%
\begin{eqnarray}
&&\frac{\left. _{2}F_{1}\right. \left( -\left( 2m+1\right) ,2m+5;\frac{5}{2}%
;\sin ^{2}\left( \alpha x\right) \right) }{\left. _{2}F_{1}\right. \left(
-2m,2m+6;\frac{7}{2};\frac{1}{2}\right) }=\frac{\left( -1\right) ^{m+1}}{20}%
\frac{\left( 2m+1\right) \left( 2m+5\right) }{\left( m+1\right) \left(
m+2\right) }  \label{r42} \\
&&\times \frac{\left[ \left( 2m+3\right) \cos \left( 2\alpha \left(
2m+3\right) x\right) -\cot \left( 2\alpha x\right) \sin \left( 2\alpha
\left( 2m+3\right) x\right) \right] }{\sin ^{2}\left( 2\alpha x\right) }\ . 
\notag
\end{eqnarray}

Besides that we present results of some integrals with the combination of
trigonometric functions of Eq. (\ref{r22}) and with the hypergeometric
function of Eq. (\ref{r25}).

1. As far as the eigenfunctions $\widetilde{\chi }_{k}$ are normalized, $%
\int_{0}^{\frac{\pi }{2\alpha }}\widetilde{\chi }_{k}^{2}dx=1$, from Eq. (%
\ref{r22}) one has%
\begin{equation*}
\int_{0}^{\frac{\pi }{2\alpha }}\left[ k\cos \left( 2\alpha kx\right) -\cot
\left( 2\alpha x\right) \sin \left( 2\alpha kx\right) \right] ^{2}dx=\frac{%
\pi }{4\alpha }\left( k^{2}-1\right) \ .
\end{equation*}%
Passing to integration over the inerval $\left( 0,\pi \right) $ by a change
of variable $2\alpha x\rightarrow x$, one gets%
\begin{equation}
\int_{0}^{\pi }\left[ k\cos \left( kx\right) -\cot \left( x\right) \sin
\left( kx\right) \right] ^{2}dx=\frac{\pi }{2}\left( k^{2}-1\right) \ .
\label{r51}
\end{equation}%
Writing the eigenfunctions in terms of the hypergeometric function, Eqs. (%
\ref{r25}), (\ref{r30}), one has%
\begin{equation*}
\int_{0}^{\frac{\pi }{2\alpha }}\sin ^{4}\left( \alpha x\right) \cos
^{4}\left( \alpha x\right) \left. _{2}F_{1}^{2}\right. \left( -n,n+4;\frac{5%
}{2};\sin ^{2}\left( \alpha x\right) \right) dx=\frac{\pi }{4\alpha }\left(
\left( n+2\right) ^{2}-1\right) C_{n}^{2}\ .
\end{equation*}%
Passing to integration over the inerval $\left( 0,\pi /2\right) $ by a
change of variable $\alpha x\rightarrow x$, one gets%
\begin{equation}
\int_{0}^{\frac{\pi }{2}}\sin ^{4}x\cos ^{4}x\left. _{2}F_{1}^{2}\right.
\left( -n,n+4;\frac{5}{2};\sin ^{2}\left( x\right) \right) dx=\frac{\pi }{4}%
\left( \left( n+2\right) ^{2}-1\right) C_{n}^{2}\ .  \label{r52}
\end{equation}%
Making a change of variable $z=\sin ^{2}x$ in Eq. (\ref{r52}), one obtains%
\begin{equation}
\int_{0}^{1}z^{\frac{3}{2}}\left( 1-z\right) ^{\frac{3}{2}}\left.
_{2}F_{1}^{2}\right. \left( -n,n+4;\frac{5}{2};z\right) dz=\frac{\pi }{2}%
\left( \left( n+2\right) ^{2}-1\right) C_{n}^{2}\ .  \label{r53}
\end{equation}%
The coefficient $C_{n}$ in Eqs. (\ref{r52}) and (\ref{r53}) must be taken in
dependence of even or odd $n$ as%
\begin{eqnarray}
C_{n}^{even} &=&\frac{\left( -1\right) ^{\frac{n}{2}+1}\left.
_{2}F_{1}\right. \left( -n,n+4;\frac{5}{2};\frac{1}{2}\right) }{4\left(
n+2\right) },\   \label{r54} \\
C_{n}^{odd} &=&\frac{\left( -1\right) ^{\frac{n-1}{2}+1}n\left( n+4\right)
\left. _{2}F_{1}\right. \left( -n+1,n+5;\frac{7}{2};\frac{1}{2}\right) }{%
20\left( \left( n+2\right) ^{2}-1\right) }\ .  \notag
\end{eqnarray}

2. One can demostrate that $\widetilde{\chi }_{k}^{2}\left( x\right) $ is
symmetric relative to the middle of the interval $x=\frac{\pi }{4\alpha }$:%
\begin{equation*}
\widetilde{\chi }_{k}^{2}\left( -\left( x-\frac{\pi }{4\alpha }\right)
\right) =\widetilde{\chi }_{k}^{2}\left( x-\frac{\pi }{4\alpha }\right) \ .
\end{equation*}%
Then one can write for the expected value of the coordinate $x$ in a state $%
\widetilde{\chi }_{n}$:%
\begin{equation}
\left\langle x\right\rangle _{k}=\int_{0}^{\frac{\pi }{2\alpha }}\widetilde{%
\chi }_{k}^{\ast }\widehat{x}\widetilde{\chi }_{k}dx=\int_{0}^{\frac{\pi }{%
2\alpha }}\widetilde{\chi }_{k}^{2}\left( x\right) xdx=\int_{-\frac{\pi }{%
4\alpha }}^{\frac{\pi }{4\alpha }}\widetilde{\chi }_{k}^{2}\left( \xi
\right) \left( \xi +\frac{\pi }{4\alpha }\right) d\xi  \label{r61}
\end{equation}%
\begin{equation*}
=\frac{\pi }{4\alpha }\int_{-\frac{\pi }{4\alpha }}^{\frac{\pi }{4\alpha }}%
\widetilde{\chi }_{k}^{2}\left( \xi \right) d\xi =\frac{\pi }{4\alpha }\ ,
\end{equation*}%
where $\xi =x-\frac{\pi }{4\alpha }$. Substituting $\widetilde{\chi }%
_{k}\left( x\right) $ from Eq. (\ref{r22}) into Eq. (\ref{r61}), one has%
\begin{equation*}
\int_{0}^{\frac{\pi }{2\alpha }}\left[ k\cos \left( 2\alpha kx\right) -\cot
\left( 2\alpha x\right) \sin \left( 2\alpha kx\right) \right] ^{2}xdx=\frac{%
\pi ^{2}}{16\alpha ^{2}}\left( k^{2}-1\right) \ .
\end{equation*}%
Passing to integration over the inerval $\left( 0,\pi \right) $ by a change
of variable $2\alpha x\rightarrow x$, one gets%
\begin{equation}
\int_{0}^{\pi }\left[ k\cos \left( kx\right) -\cot \left( x\right) \sin
\left( kx\right) \right] ^{2}xdx=\frac{\pi ^{2}}{4}\left( k^{2}-1\right) \ .
\label{r62}
\end{equation}%
For the eigenfunctions in terms of the hypergeometric function, Eqs. (\ref%
{r25}), (\ref{r30}), one has%
\begin{equation*}
\int_{0}^{\frac{\pi }{2\alpha }}\sin ^{4}\left( \alpha x\right) \cos
^{4}\left( \alpha x\right) \left. _{2}F_{1}^{2}\right. \left( -n,n+4;\frac{5%
}{2};\sin ^{2}\left( \alpha x\right) \right) xdx=\frac{\pi ^{2}}{16\alpha
^{2}}\left( \left( n+2\right) ^{2}-1\right) C_{n}^{2}
\end{equation*}%
Passing to integration over the inerval $\left( 0,\pi /2\right) $ by a
change of variable $\alpha x\rightarrow x$, one gets%
\begin{equation}
\int_{0}^{\frac{\pi }{2}}\sin ^{4}\left( x\right) \cos ^{4}\left( x\right)
\left. _{2}F_{1}^{2}\right. \left( -n,n+4;\frac{5}{2};\sin ^{2}\left(
x\right) \right) xdx=\frac{\pi ^{2}}{16}\left( \left( n+2\right)
^{2}-1\right) C_{n}^{2}\ .  \label{r63}
\end{equation}%
The coefficient $C_{n}$ in Eq. (\ref{r63}) must be taken in dependence of
even or odd $n$ as is given in Eq. (\ref{r54}).

\end{document}